\documentclass[conference,letterpaper]{IEEEtran}

\usepackage{cite}
\usepackage{bm}
\usepackage{amsmath}
\usepackage{extarrows}
\usepackage{amssymb}
\usepackage{graphicx}
\usepackage{color}
\usepackage{enumerate}
\usepackage{bookmark} 
\graphicspath{{figure}}
\usepackage{setspace}

\newcommand{\mr}{\mathrm}

\newcommand{\BE}{\begin{equation}}
\newcommand{\EE}{\end{equation}}
\newcommand{\BS}{\begin{subequations}}
\newcommand{\ES}{\end{subequations}}
\renewcommand{\bf}{\bm}

\newtheorem{theorem}{Theorem}

\newtheorem{assumption}{Assumption}

\newtheorem{lemma}{Lemma}

\ifCLASSINFOpdf
\graphicspath{{figure/}}
\else
\fi
\begin{document}

\title{{Capacity Optimality of OAMP in Coded Large Unitarily Invariant Systems}}

\author{\IEEEauthorblockN{Lei~Liu\IEEEauthorrefmark{1}\IEEEauthorrefmark{2},  \emph{Member, IEEE},  Shansuo~Liang\IEEEauthorrefmark{1}\IEEEauthorrefmark{3},  and~Li~Ping\IEEEauthorrefmark{1}, \emph{Fellow, IEEE}}\vspace{-3mm}\\\normalsize
           \IEEEauthorrefmark{1}Department of Electronic Engineering, City University of Hong Kong, Hong Kong\\
           \IEEEauthorrefmark{2}School of Information Science, Japan Institute of Science and Technology (JAIST), Nomi 923-1292, Japan\\
           \IEEEauthorrefmark{3}Theory lab, Central Research Institute, 2012 Labs, Huawei Technologies Co., Ltd.
           }

\maketitle

\begin{abstract}
This paper investigates a large unitarily invariant system (LUIS) involving a unitarily invariant sensing matrix, an arbitrary fixed signal distribution, and forward error control (FEC) coding. Several area properties are established based on the state evolution of orthogonal approximate message passing (OAMP) in an un-coded LUIS.  Under the assumptions that the state evolution for joint OAMP and FEC decoding is correct and the replica method is reliable, we  analyze the achievable rate of OAMP. We prove that OAMP reaches the constrained capacity predicted by the replica method of the LUIS with an arbitrary signal distribution based on matched FEC coding. Meanwhile, we elaborate a constrained capacity-achieving coding principle for LUIS, based on which irregular low-density parity-check (LDPC) codes are optimized for binary signaling in the simulation results. We show that OAMP with the optimized codes has significant performance improvement over the un-optimized ones and the well-known Turbo linear MMSE algorithm. For quadrature phase-shift keying (QPSK) modulation,  constrained capacity-approaching bit error rate (BER) performances are observed under various  channel conditions. 

\end{abstract}

\textit{ A full version of this paper is accessible at
\href{https://arxiv.org/pdf/2108.08503.pdf}{arXiv} (see \cite{Lei_cap_OAMP}).} 

\section{Introduction}
 
Consider estimating $\bf{x}=\{x_i\}\in\mathbb{C}^{N}$ from its observation $\bf{y}\in\mathbb{C}^{M}$ in a linear system\vspace{-1mm}
\BE\label{Eqn:linear_system} 
\bf{y}=\bf{Ax}+\bf{n},\vspace{-1mm}
\EE
where $\bf{n}\in \mathbb{C}^{M}$ contains  Gaussian noise samples. For convenience, we refer to $\bf{A}\in\mathbb{C}^{M\times N}$ as a sensing matrix. We assume that $\{x_i\}$ follow a fixed distribution $ x_i \sim P_X,\forall i$. Furthermore, we assume that $\bf{x}$ is generated by a forward error control (FEC) code that includes un-coded $\bf{x}$ as a special  case. 

A wide range of communication applications can be modeled by \eqref{Eqn:linear_system}. A classic one is a multiple-input multiple-output (MIMO) system where $\bf{A}$ is a channel coefficient matrix \cite{Biglieri2007,David2005} and $P_X$ is determined by the signaling (i.e., modulation) method. Continuous Gaussian signaling, with which $\{x_i\}$ are independently and identically distributed (IID) Gaussian (IIDG), is often assumed in information theoretical studies but, due to implementation concerns, all practically used signaling methods are discrete such as  quadrature phase-shift keying (QPSK). 

A more recent application of \eqref{Eqn:linear_system} is the so-called massive-access scheme where $\bf{A}$ consists of pilot signals  and $P_X$ is jointly determined by channel coefficients and user activity \cite{Yuwei2018, Yuwei20182}. In this case, a common assumption of $P_X$ is Bernoulli-Gaussian.  Massive-access has attracted wide research interests for machine-type communications in the $5^{\rm th}$ and $6^{\rm th}$ generation (5G and 6G) cellular systems. 

Two performance measures are commonly used for the system in \eqref{Eqn:linear_system}, namely mean squared error (MSE) for an un-coded system and achievable rate for a FEC coded one. The optimal limits in these two cases are respectively given by minimum MSE (MMSE) and information theoretic capacity. For simplicity, we say that a receiver is MMSE-optimal if its MSE achieves MMSE for un-coded $\bf{x}$, or capacity-optimal if its achievable rate reaches mutual information $I(\bf{x}; \bf{y})$ 
for coded $\bf{x}$. Optimal receivers under both measures generally have prohibitively high complexity \cite{Micciancio2001,verdu1984_1} except for a few cases listed below. 
\begin{itemize}
  \item The classic linear MMSE (LMMSE) detector is optimal when \eqref{Eqn:linear_system} is un-coded with IIDG signaling \cite{Kay1993}.   
  \item Some compressed-sensing algorithms are asymptotically MMSE-optimal when $\bf{x}$ is un-coded and sparse. 
  \item When $\bf{A}$ is diagonal, \eqref{Eqn:linear_system} is equivalent to a set of parallel single-input-single-output (SISO) sub-systems. Turbo \cite{Berrou1993, douillard1995iterative} or low-density parity-check (LDPC) \cite{Gallager1962, Chung01} codes are nearly capacity-optimal in this case.
  \item For Gaussian signaling, Turbo-type detection algorithms \cite{Wang1999} are capacity-optimal using proper coding \cite{Yuan2014, Lei20161b,YC2018TWC}. 
\end{itemize}
For discrete signaling with dense $\bf{A}$ and $\bf{x}$, however, detection with proven MMSE/capacity optimality and practical complexity remained a difficult issue until recently \cite{Barbier2018b, Kabashima2006, Tulino2013, Reeves_TIT2019, Barbier2017arxiv,  LeiTIT}. 

Approximate message passing (AMP) represents a remarkable progress measured by both types of optimality with practical complexity. AMP employs a so-called Onsager term to regulate the correlation problem in iterative processing \cite{Donoho2009}. A distinguished feature of AMP is that its asymptotic MSE performance can be accurately analyzed by a state-evolution (SE) technique \cite{Bayati2011}. Based on SE, the MMSE optimality of AMP is proved in \cite{Reeves_TIT2019, Barbier2017arxiv}. The capacity optimality of AMP is proved in \cite{LeiTIT}.  

We will say that $\bf{A}$ in \eqref{Eqn:linear_system} is IIDG when its entries are IIDG. Good performance of AMP is guaranteed only when $\bf{A}$ is IIDG. This IIDG restriction is relaxed in orthogonal AMP (OAMP) \cite{Ma2016,MaSPL2015a, Ma_SPL2015b}. Let the singular value decomposition (SVD) of $\bf{A}$ be $\bf{A}\!=\!\bf{U}^{\rm H}\bf{\Sigma} \bf{V}$,  where $\bf{U}\!\in\! \mathbb{C}^{M\times M}$ and $\bf{V}\!\in\! \mathbb{C}^{N\times N}$ are unitary  and $\bf{\Sigma}$ is an $M\times N$ rectangular diagonal matrix. We will say that $\bf{V}$ is Haar distributed if $\bf{V}$ is uniformly distributed over all unitary matrices \cite{Hiai2000, Tulino2004}.  We will say that $\bf{A}$ is \emph{right-unitarily invariant} if $\bf{V}$ is Haar distributed. We will call \eqref{Eqn:linear_system} a large unitarily invariant system (LUIS) when $M, N\to\infty$ with their ratio fixed and $\bf{V}$ is Haar distributed. Recently, to avoid the high complexity LMMSE in OAMP, low-complexity convolutional AMP (CAMP) \cite{Takeuchi2020CAMP} and memory AMP (MAMP) \cite{LeiMAMP} were proposed for LUIS.  The MMSE and constrained capacity of LUIS can be predicted by the replica method from statistical physics \cite{Tulino2013, Kabashima2006}. However, the replica method involves an exchange of limits and the assumption of replica symmetry, which are unproven for general LUIS. Recently, it is proven that the MMSE and constrained capacity predicted by the replica method are correct for a sub-class of LUIS \cite{Barbier2018b}. For more cases, a rigorous proof of the replica method is still open issues. 

This paper concerns the capacity optimality. One of the tools used this paper is  the SE technique for OAMP that is conjectured for LUIS in \cite{Ma2016} and rigorously proved in \cite{Rangan2016, Takeuchi2017}. Another tool is the replica method is a major tool in our derivation. For convenience, we will call the MMSE and constrained capacity predicted by the replica method as ``replica MMSE" and ``replica constrained capacity", respectively. We will say that a receiver is MMSE-optimal if its MSE achieves the replica MMSE when $\bf{x}$ is un-coded, or capacity-optimal if its achievable rate reaches the replica constrained capacity when $\bf{x}$ is coded. We will show that OAMP can achieve the constrained capacity of LUIS provided that the SE and replica methods are reliable for OAMP. 

As related works,  the MMSE-optimality of OAMP in LUIS was derived in \cite{Rangan2016, Takeuchi2017} under the assumption that the SE for OAMP is accurate. We demonstrated the capacity optimality of AMP in \cite{LeiTIT} using the I-MMSE and area properties \cite{Guo2005, Bhattad2007}. As mentioned above, AMP is most suited to systems with IIDG sensing matrices. Unitarily-invariant matrices form a larger set that includes IIDG matrices as a special case. Hence OAMP is applicable to a wider range of applications than AMP. The extension of the derivations in \cite{LeiTIT} from AMP to OAMP is not straightforward, since orthogonal local estimators are generally not (locally) MMSE-optimal, so the I-MMSE property cannot be used directly in OAMP. In this paper, we will circumvent this difficulty  by properly remodeling the OAMP receiver structure.



AMP-type algorithms, including OAMP, all involve iteration between two local processors such as linear estimator (LE) $\gamma$ and non-linear estimator (NLE) $\phi$  \cite{Donoho2009,Ma2016,Rangan2016}. The performance of these two local processors can be respectively characterized by two transfer functions  $\gamma_{\rm SE}$ and $\phi_{\rm SE}$ using the SE technique \cite{Bayati2011,Takeuchi2017, Rangan2016}.   Following \cite{LeiTIT}, we call mutual information $I(\bf{x}; \bf{y})$ the constrained capacity of the system in \eqref{Eqn:linear_system}. We proved the capacity optimality of AMP in \cite{LeiTIT} based on the assumption of MMSE-optimal NLE $\phi$. 

Unfortunately, we cannot assume the local processors in OAMP are MMSE-optimal. This is due to the requirement of input-output error orthogonality on the local processors in OAMP. MMSE-optimal processors are generally not orthogonal, so the local processors in OAMP are generally not MMSE-optimal. This constitutes a main difficulty in extending the results on AMP in \cite{LeiTIT} to OAMP. In this paper, we  overcome the difficulty using  an equivalent structure of OAMP such that the NLE  can be MMSE-optimal. It allows the use of the mutual information-MMSE (I-MMSE) theorem to obtain the achievable rate of OAMP. We prove that this achievable rate is equal to the replica constrained capacity, which leads to the  capacity optimality of OAMP. The proof of  capacity optimality of OAMP in this paper can also be extended to the vector AMP algorithm (VAMP) \cite{Rangan2016}  and expectation propagation (EP) \cite{Cakmak2018, opper2005expectation, Minka2001} due to the algorithmic equivalence between these algorithms. Such equivalence is noted in \cite{MaTWC, Takeuchi2020CAMP, Takeuchi2017}. 

OAMP is being extensively investigated in many emerging applications including MIMO channel estimation and massive access \cite{MaTWC, Yiyao2021, Khani2020, Zhang2019} since it is not restricted to systems with Gaussian signaling, sparsity and IIDG sensing matrices. Most exiting works on OAMP are for un-coded case. The findings in this paper provide an optimization technique for coded cases. 
    

\section{System Model and Preliminaries}%

A large unitarily invariant system (LUIS) in \eqref{Eqn:linear_system} consists of a linear constraint $\Gamma$ and a non-linear constraint $\Phi$:
\BS\begin{align}
{\rm Linear\; constraint}\;\; \Gamma:\quad &\bf{y}=\bf{Ax}+\bf{n},  \label{Eqn:LC}\\
{\rm Non\!\!-\!\!linear\; constraint}\;\; \Phi:\quad  &\bf{x}\sim P_{\bf{X}}(\bf{x}), \label{Eqn:NLC} 
\end{align}\ES
 where $\bm{y}\!\in\!\mathbb{C}^{M\!\times\!1}$ is an observed vector, $\bf{A}\!\in\!\mathbb{C}^{M\!\times\! N}$ a \emph{right-unitarily invariant} measurement matrix and $\bm{n}\!\sim\!\mathcal{CN}(\mathbf{0},\sigma^2\bm{I}_M)$ a Gaussian noise vector.  The average power of  $\bf{x}$ is normalized, i.e., $\frac{1}{N}\mr{E}\{\|{\bf{x}}\|^2\}=1$, and  $snr = \sigma^{-2}$ is the transmit SNR. We consider a large-scale LUIS that $M,N\to\infty$ with a fixed $\beta=N/M$. Only the receiver knows $\bf{A}$\footnote{If the transmitter also has $\bf{A}$,  the LUIS can be converted to parallel SISO  channels. Then, water filling is capacity optimal.}.  This assumption was frequently used in multiple-input-multiple-output (MIMO)  communications \cite{MaTWC, Lei20161b}.
 

For an un-coded LUIS, the constraint  \eqref{Eqn:NLC} can be rewritten in a symbol-by-symbol manner as:
\BE\label{Eqn:Phi_S}
    {\rm Constellation\; constraint}\;\; \Phi_{\cal S}:\quad x_i\sim P_X(x),\; \forall i.
\EE
That is, the entries of $\bf{x}$ are IID with distribution $P_X(x)$.  

In an un-coded LUIS, mean squared error (MSE) is commonly used for performance measurement. If $P_X(x)$ is a Gaussian distribution, the optimal solution is the standard linear minimum MSE (MMSE) estimate. For non-Gaussian distribution $P_X(x)$, finding the optimal solution is generally NP hard \cite{Micciancio2001,verdu1984_1}.  
   

In an un-coded LUIS, an error-free estimation can not be guaranteed. To ensure an error-free estimation, we consider the LUIS with forward error control (FEC) coding. For a coded LUIS, the  constraint in \eqref{Eqn:NLC} is rewritten to a code constraint:
\BE\label{Eqn:Phi_C}
   {\rm Code\; constraint}\;\; \Phi_{\cal C}:\quad \bf{x} \in \bf{\mathcal{C}} \;\; {\rm and} \;\; x_i\sim P_X(x),\; \forall i,
\EE
where the entries of $\bf{x}$ have the identical distribution $P_X(x)$. 

In a coded LUIS, achievable rate is commonly used for performance metrics. For a specific receiver, the code rate $R_{\bf{\mathcal{C}}}$ of codebook $\bf{\mathcal{C}}$ is an achievable rate, if its estimation of $\bf{x}$ under code constraint $\bf{\mathcal{C}}$ is error-free, i.e., ${\rm mmse}\{\bf{x}|\bf{y}, \bf{A}, {\Gamma}, \Phi_{\cal C} \}\equiv \tfrac{1}{N}{\mr{E}}\{\|{\rm E}\{\bf{x}|\bf{y}, \bf{A}, {\Gamma}, \Phi_{\cal C} \}-{\bf{x}}\|^2\}\to 0$ or equivalently, the block error probability$\to 0$. A natural upper bound of achievable rate is the constrained capacity of LUIS, i.e., i.e., the mutual information given $x\sim P_X(x)$:
\BE
   R_{\bf{\mathcal{C}}}\le C_{\rm LUIS}=I(\bf{x};\bf{y}).
\EE
 
 Our aim is to find a coding scheme and a receiver with practical complexity such that the achievable rate approaches the constrained capacity of LUIS, i.e., $ R_{\bf{\mathcal{C}}}\to C_{\rm LUIS}$. 

 For arbitrary signal distributions, the constrained capacity of LUIS is not trivial and was conjectured by replica method for right-unitarily-invariant matrices in \cite{Tulino2013, Kabashima2006}.  

\begin{lemma} [Constrained Capacity] \label{Pro:dis_cap}
 Suppose that $\rho^*$ is the unique solution of $\rho^* = snr\,\mathcal{R}_{\bf{R}}\big(-snr\,v^*\big)$ where $v^*={\rm mmse}\{\bf{x}|\sqrt{\rho^*}\bf{x}+\bf{z},\Phi\}$  denotes the replica MMSE of a LUIS \cite{Tulino2013, Kabashima2006, Barbier2018b}. The constrained capacity of a LUIS predicted by replica  method \cite{Tulino2013, Kabashima2006, Barbier2018b} is given by \vspace{-0.2cm}
\BE\label{Eqn:dis_cap_ref}
    C^{\rm Rep}_{\rm LUIS}(snr)\! =\!\!\! \int_0^{v^*\!snr} \!\!\! \mathcal{R}_{\bf{R}}(-z)dz + I(x;\sqrt{\rho^*}x+z)  - \rho^*v^*,
\EE
where $z\sim\mathcal{CN}(0,1)$, and $\mathcal{R}_{\bf{R}}(\cdot)$ is the \emph{R-transform} of matrix $\bf{R}=\bf{A}^H\bf{A}$ (see \cite{Tulino2004} for an introduction to this transform).
\end{lemma}

Since the replica method is heuristic, the exact constrained capacity of LUIS is still an open issue. Recently, the replica results were justified in \cite{Barbier2018b} for a sub-class of right-unitarily-invariant matrices $\bf{A} = \bf{U} \bf{W}$,
where $\bf{W}$ is IIDG, $\bf{U}$ is a product of a finite number of independent matrices, each with IID matrix-elements that are either bounded or standard Gaussian, and $\bf{U}$ and $\bf{W}$ are independent.   

How to design a coding scheme and a receiver with practical complexity to achieve  the replica constrained capacity is still an open issue. In this paper, we will prove the  capacity optimality of OAMP based on a matching principle.

\begin{figure}[b!]
  \centering
  \includegraphics[width=4.5cm]{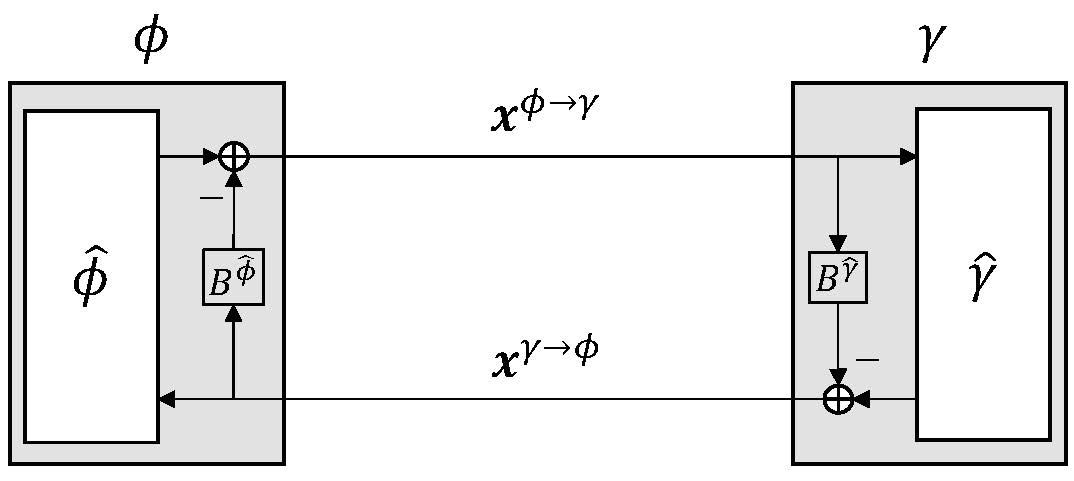}\\\vspace{-0.1cm}
  \caption{Graphical illustration for OAMP receiver involving two local orthogonal estimators $\gamma$ (for $\Gamma$) and $\phi$ (for $\Phi$) , which are constructed by GSO based on the prototypes $\hat{\gamma}$ and $\hat{\phi}$, respectively.}\label{Fig:OAMP} \vspace{-0.2cm}
\end{figure}  

\section{Orthogonal AMP (OAMP)}\label{Sec:OAMP}
In this paper, we consider locally optimal prototypes:
\BS\label{Eqn:MMSE_funs}\begin{align}
  {\rm  LE:} \quad  \hat{\gamma}_t\big(\bf{x}^{\phi\to \gamma}_{t}\big) &={\rm E}\{\bf{x}|\bf{x}^{\phi\to \gamma}_{t},\Gamma\},\\
  {\rm NLE:} \quad   \hat{\phi}_t\big(\bf{x}^{\gamma\to \phi}_t \big) &={\rm E}\{\bf{x}|\bf{x}^{\gamma\to \phi}_t,\Phi\}.
\end{align}\ES 
Let the messages be in Gram-Schmidt (GS) models \cite{Lei_cap_OAMP}: 
\BE\label{Eqn:GSO_gamma_phi}  
  \bf{x}^{\phi\to \gamma}_{t} = \alpha_{t}^{\phi\to \gamma}\bf{x}+\bf{\xi}_{t}^{\phi\to \gamma}, \quad \bf{x}^{\gamma\to \phi}_t   = \alpha_{t}^{\gamma\to \phi}\bf{x}+{\bf{\xi}}^{\gamma\to \phi}_{t}.
 \EE
Let the average powers of the GS errors $\bf{\xi}_{t}^{\phi\to \gamma}$ and ${\bf{\xi}}^{\gamma\to \phi}_{t}$ be $v^{\phi\to \gamma}$ and $v^{\gamma\to \phi}_{t}$, respectively.  OAMP can be constructed as
\BS\label{Eqn:OAMP}\begin{align}
     \bf{x}^{\gamma\to \phi}_t & = {\gamma}_t\big(\bf{x}^{\phi\to \gamma}_{t}\big)= \hat{\gamma}_t\big(\bf{x}^{\phi\to \gamma}_{t}\big)- B_{\hat{\gamma}_t} \bf{x}^{\phi\to \gamma}_{t}, \\
      \bf{x}^{\phi\to \gamma}_{t+1} &= {\phi}_t\big(\bf{x}^{\gamma\to \phi}_t \big) = \hat{\phi}_t\big(\bf{x}^{\gamma\to \phi}_t \big)- B_{\hat{\phi}_t} \bf{x}^{\gamma\to \phi}_{t},
\end{align}\ES
where $B_{\hat{\gamma}_t}$ and $B_{\hat{\phi}_t}$ are GS coefficients given by
\BS\label{Eqn:GSO_B}\begin{align}
B_{\hat{\gamma}_t} = \mr{E}\{(\bf{\xi}^{\phi\to \gamma}_{t})^{\mr{H}}\hat{\gamma}_t\big(\bf{x}^{\phi\to \gamma}_{t}\big)\}/\mr{E}\{\|\bf{\xi}^{\phi\to \gamma}_{t}\|^2\},\\ 
B_{\hat{\phi}_t} = \mr{E}\{(\bf{\xi}^{\gamma\to \phi}_{t})^{\mr{H}}\hat{\phi}_t\big(\bf{x}^{\gamma\to \phi}_{t}\big)\}/\mr{E}\{\|\bf{\xi}^{\gamma\to \phi}_{t}\|^2\}.
\end{align}
\ES 
Refer to \cite{Lei_cap_OAMP} for specific calculations of the GS coefficients.  Fig.~\ref{Fig:OAMP} gives the block diagram of OAMP.

 We define the true errors as $\bf{e}_{t}^{{\phi}\to\gamma}  \equiv  \bf{x}_{t}^{{\phi}\to\gamma}  -\bf{x}$, $\bf{e}_{t}^{{\gamma}\to\phi}  \equiv  \bf{x}_{t}^{{\gamma}\to\phi}  -\bf{x}$, $ v_t   \equiv  \tfrac{1}{N} {\rm E}\big\{\| \bf{e}_{t}^{{\phi}\to\gamma}   \|^2\big\}$ and $\rho_t   \equiv N/ {\rm E}\big\{\| \bf{e}_{t}^{\gamma\to{\phi}} \|^2\}$. The following lemma was proved in \cite{Rangan2016, Takeuchi2017} for OAMP in an un-coded LUIS with IID input $\bf{x}$. 

\begin{lemma} 
[Asymptotic IIDG]\label{The:SE}
Let $\bf{x}\sim \Phi_{\cal S}$ (see \eqref{Eqn:Phi_S}). In OAMP, $\bf{e}_{t}^{\gamma\to{\phi}}$ can be modeled by a sequence of IIDG samples independent of $\bf{x}$. The $\gamma_t$ and ${\phi}_t$ in  OAMP can be characterized by the following transfer functions:
\BS\label{Eqn:TF1}\begin{align}
&\rho = {\gamma}_{\rm SE}(v)\equiv [\hat{\gamma}_{\rm SE}(v)]^{-1} - v^{-1},\\   
&v   = {\phi}_{\rm SE}^{\cal S}(\rho)\equiv \big([\hat{\phi}_{\rm SE}^{\cal S}(\rho)]^{-1}- \rho \big)^{-1}.
\end{align}\ES
where $ \hat{\gamma}_{\rm SE}(v) = \tfrac{1}{N}{\mr{tr}}\left\{ [snr\bf{A}^H\bf{A} + v^{-1}\,\bf{I}]^{-1}\right\} $, $ \hat{\phi}_{\rm SE}^{\cal S}(\rho) ={\rm mmse}\{\bf{x}|\sqrt{\rho}\bf{x}+\bf{z},\Phi_{\cal S}\}$,  $\bf{z}\sim\mathcal{CN}(\bf{0},\bf{I})$  and $\hat{\gamma}_{\rm SE}^{-1}(\cdot)$ is the inverse of $\hat{\gamma}_{\rm SE}(\cdot)$.
\end{lemma}

 \section{Area Properties of LUIS} 
 
 \begin{figure}[b!] 
  \centering
  \includegraphics[width=3.5cm]{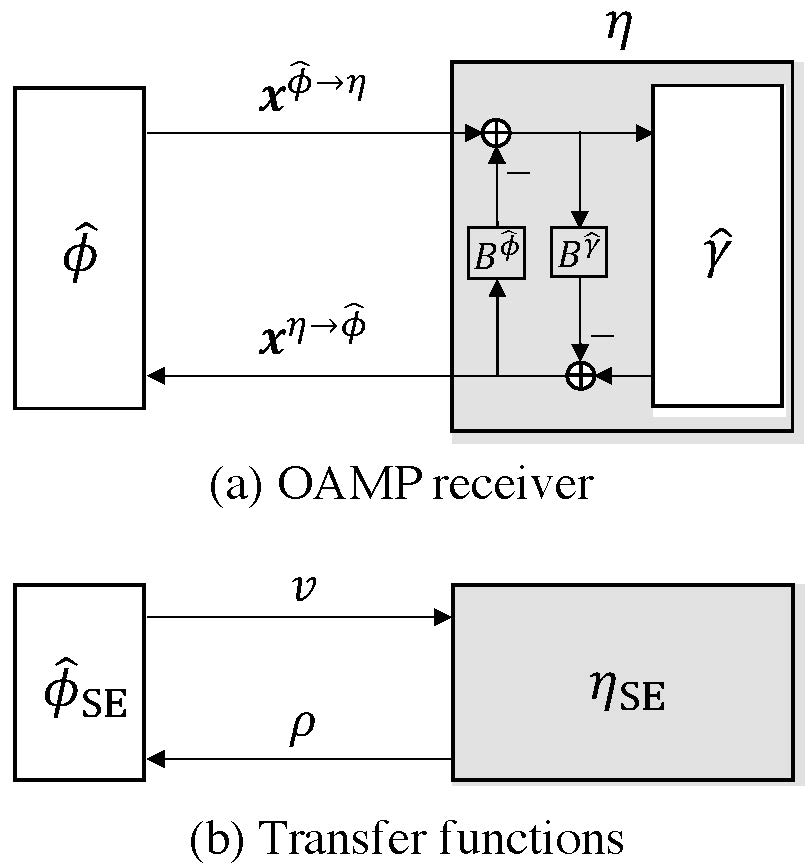}\\\vspace{-0.1cm}
  \caption{Graphical illustrations of  OAMP receiver and the transfer functions.}\label{Fig:SE_OAMP} \vspace{-0.2cm}
\end{figure} 
The I-MMSE property derived in \cite{Guo2005, Bhattad2007} establishes a connection between achievable rate and MMSE performance. Below, we will apply this property to examine the capacity optimality of OAMP.  In general, an MMSE-optimal estimator does not meet the orthogonal requirement defined in Section \ref{Sec:OAMP}. This constitutes a difficulty in directly applying the I-MMSE property to OAMP. In this subsection, we outline an equivalent structure of OAMP, which circumvents this difficulty. We rewrite the OAMP in \eqref{Eqn:OAMP} to
 \begin{align}\label{OAMP_simple}
    \bf{x}_{t}^{\eta\to\hat{\phi}} = \eta_t\big(\bf{x}_{t}^{\hat{\phi}\to\eta}\big) \quad {\rm and} \quad 
   \bf{x}_{t+1}^{\hat{\phi}\to\eta}  = \hat{\phi}_t\big(\bf{x}_{t}^{\eta\to\hat{\phi}}\big), 
\end{align} 
where $\eta_t$ includes $\hat{\gamma}_t$ and the GSO operations.  Fig.~\ref{Fig:SE_OAMP}(a) gives a graphical illustration of \eqref{OAMP_simple}. 

 Let $\rho_t   \equiv {{N}/{\rm E}\big\{\| \bf{x}_{t}^{\eta\to\hat{\phi}} -\bf{x} \|^2\}}$ and $v_t   \equiv  \tfrac{1}{N} {\rm E}\big\{\| \bf{x}_{t}^{\hat{\phi}\to\eta} -\bf{x}   \|^2\big\}$. We then rewrite the transfer functions in \eqref{Eqn:TF1} to
 \begin{align}\label{Eqn:TF}
\rho  = {\eta}_{\rm SE}(v)\equiv v^{-1}-[\hat{\gamma}_{\rm SE}^{-1}(v)]^{-1} \quad {\rm and} \quad 
v   = \hat{\phi}_{\rm SE}^{\cal S}(\rho).
\end{align} 
 The following properties are easy to verify.
\begin{itemize}
    \item The fixed-point equation of $\hat{\phi}_{\rm SE}^{\cal S}(\rho)= {\eta}^{-1}_{\rm SE}(\rho)$ is the same as that of ${\phi}_{\rm SE}^{\cal S}(\rho)= {\gamma}^{-1}_{\rm SE}(\rho)$, where ${\eta}^{-1}_{\rm SE}(\cdot)$ and ${\gamma}^{-1}_{\rm SE}(\cdot)$ are the inverses of ${\eta}_{\rm SE}(\cdot)$ and ${\gamma}_{\rm SE}(\cdot)$, respectively.
    \item ${\eta}^{-1}_{\rm SE}(\rho)< \hat{\phi}_{\rm SE}^{\cal S}(\rho)$ if and only if ${\gamma}^{-1}_{\rm SE}(\rho)< {\phi}_{\rm SE}^{\cal S}(\rho)$, for $\rho \le \rho^*$, where $\rho^*$ is the fixed point of $\hat{\phi}_{\rm SE}^{\cal S}(\rho)= {\eta}^{-1}_{\rm SE}(\rho)$.
\end{itemize}

\begin{figure}[t] 
  \centering
  \includegraphics[width=5.5cm]{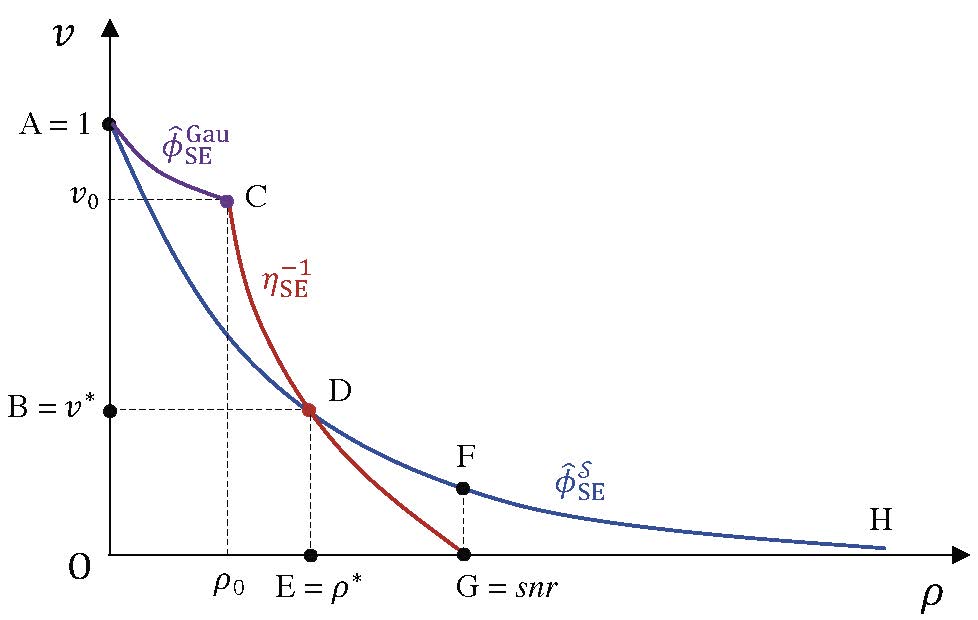}\\ 
  \caption{Graphic illustration of OAMP in the un-coded LUIS. $(\rho^*, v^*)$ is the fixed point of the transfer functions in \eqref{Eqn:TF}. ${\rm H} = (0, \infty)$, $\hat{\phi}_{\rm SE}^{\rm Gau}(\rho)=1/(1+\rho)$, $\rho_0=[\gamma_{\rm SE}(1)]^{-1}-1$, $v_0=\gamma^{-1}_{\rm SE}(\rho_0)$ and $\eta_{\rm SE}(0)=snr$.  } \label{Fig:TF_chart} 
\end{figure}

Fig. \ref{Fig:TF_chart} provides a graphical illustration of  \eqref{Eqn:TF1}.




 \begin{assumption}\label{Pro:SCP}
There is exactly one fixed point for $\hat{\phi}_{\rm SE}^{\cal S}(\rho) = {\eta}^{-1}_{\rm SE}(\rho)$ in $\rho>0$, where  ${\eta}^{-1}_{\rm SE}(\cdot)$ is the inverse of ${\eta}_{\rm SE}(\cdot)$.
\end{assumption}

It is proved that the SE of OAMP is convergent \cite{Takeuchi2021OAMP, Lei_SSMAMP}. Let  $(\rho^*, v^*)$ with $v^*=\hat{\phi}_{\rm SE}^{\cal S} (\rho^*)$ be the unique fixed point. 

\begin{theorem}\label{The:area_LUIS}
 Suppose that Assumption \ref{Pro:SCP} holds. Then the replica constrained capacity of a LUIS given $p_X(x)$ is given by the area $A_{\rm ADGO}$ in Fig. \ref{Fig:TF_chart} covered by ${\eta}_{\rm SE}^{-1}$ and $\hat{\phi}_{\rm SE}^{\cal S}$, i.e., 
\BS\label{Eqn:dis_cap}\begin{align}
   \!\!\!  C^{\rm Rep}_{\rm LUIS} &= A_{\rm ADGO} = \!\!\! \int_{0}^{\rho^*}\!\! \!\!\!{\hat{\phi}_{\rm SE}^{\cal S}(\rho) \,d\rho} +\log  v^* \!\!+\!  { \tfrac{1}{N} \log | \bf{B} |},  
\end{align} 
\ES
 where  $\bf{B}= \big([v^*]^{-1} - \rho^*\big)\bf{I} + snr\bf{A}^H\bf{A}$.  
\end{theorem} 
 
We have the area properties below as illustrated in Fig.~\ref{Fig:TF_chart}. 
\begin{itemize}

\item  Area $A_{\rm BDEO}$ is given by $ A_{\rm BDEO}=\rho^*v^*$, where $(\rho^*, v^*)$ denotes the unique fixed point D in Fig.~\ref{Fig:TF_chart}.

\item Area $A_{\rm AHO}$ equals to the entropy of the constellation, i.e.,  $A_{\rm AHO} = \log|\mathcal{S}|$,
i.e.,  the maximum achievable rate of a noiseless linear system in \eqref{Eqn:linear_system}. See Lemma \ref{Lem:R-MMSE}. 

\item Area $A_{\rm ADGO}$ equals to the replica constrained capacity $C^{\rm Rep}_{\rm LUIS}$ of a LUIS. See Theorem \ref{The:area_LUIS}.

\item Area $A_{\rm ACGO}$ equals to the Gaussian capacity $C_{\rm Gau}$ of a LUIS, i.e., $ A_{\rm ACGO} = C_{\rm Gau} = \tfrac{1}{N}\log \det\big(\bf{I} +  snr\bf{A}^H\bf{A}\big)$.

\item Area $A_{\rm ACD}$ represents the shaping gain of Gaussian signaling. Curve AC is the Gaussian un-coded NLE.

\item  Area $A_{\rm ADEO}$ equals to the achievable rate of a cascading receiver with OAMP detection  and decoding.  That is, $ A_{\rm ADEO} = R_{\rm CAS}$. There is no iteration between the OAMP detector and the decoder. Hence, the cascading receiver is not capacity optimal.  

\item Area $A_{\rm DGE}$ represents the rate loss for the cascading receiver, i.e., $ A_{\mr{DGE}}  = C^{\rm Rep}_{\rm LUIS}(snr) - R_{\rm CAS}(snr)$.

\item Area $A_{\rm AFGO}$ equals to the constrained capacity of a SISO channel, i.e., $ A_{\rm AFGO} = C_{\rm SISO}(snr)$.
 
\item  Area $A_{\rm DFG}$ represents the capacity gap of parallel SISO channels and a LUIS, i.e.,i.e., $A_{\rm DFG}  = C_{\rm SISO}(snr) - C^{\rm Rep}_{\rm LUIS}(snr)$. In other words, area $A_{\rm DFG}$ represents the rate loss due to the cross-symbol interference in $\bf{Ax}$. 
  
\item  Area $A_{\rm FHG}$ represents the rate loss due to the channel noise $\bf{n}$, i.e., $ A_{\rm FHG}  = A_{\rm AHO}  - A_{\rm AFGO} $. 

\item Following \eqref{Eqn:dis_cap}, we rewrite the replica constrained capacity as $C^{\rm Rep}_{\rm LUIS} \!=\!A_{\rm BDGO}\!+\! A_{\rm ADEO}\! - \!A_{\rm BDEO}$.  Since $A_{\rm BDEO}\!=\!\rho^*v^*$ and $A_{\rm ADEO}\!=\! C_{\rm SISO}(\rho^*)\!=\! I(x;\sqrt{\rho^*}x+z)$, following \eqref{Eqn:dis_cap_ref},   $ \int_0^{snr\,v^*} \mathcal{R}_{\bf{R}}(-z)dz \!=\! A_{\rm BDGO}$.
\end{itemize}

\section{Achievable Rates of OAMP in Coded LUIS}\label{Sec:rate_OAMP}
For a LUIS with FEC coding, we rewrite the problem as
\BS\begin{align}
 {\rm Linear\; constraint}\;\; \Gamma: \quad& \bf{y}=\bf{Ax}+\bf{n},\\
{\rm Code\; constraint}\;\;  \Phi_{\cal C}: \quad& \bf{x}\in  \bf{\mathcal{C}}, \;\; x_i\sim P_X(x),\; \forall i,\label{Eqn:code}
\end{align}\ES
where $\bf{\mathcal{C}}$ is a codebook. We focus on a joint OAMP and \emph{a-posteriori} probability (APP) decoding scheme for a coded LUIS. 
\BE\label{OAMP_coded} 
    \bf{x}_{t}^{\eta\to\hat{\phi}_{\cal C}}  = \eta_t\big(\bf{x}_{t}^{\hat{\phi}_{\cal C}\to\eta}\big),\qquad
   \bf{x}_{t+1}^{\hat{\phi}_{\cal C}\to\eta}  = \hat{\phi}_{t}^{\cal C}\big(\bf{x}_{t}^{\eta\to\hat{\phi}_{\cal C}}\big),
\EE
where $\hat{\phi}_{t}^{\cal C}$ is an APP decoder for the code constraint $\bf{x}\in  \bf{\mathcal{C}}$, and $\eta_t$ is the same as that in \eqref{OAMP_simple}.  Define 
\BS\label{Eqn:errors_codes}\begin{align}
   \bf{e}^{\hat{\phi}_{\cal C}\to\eta}_t    &\equiv \bf{x}^{\hat{\phi}_{\cal C}\to\eta}_t -\bf{x}, &  \rho_t &\equiv \big[{\tfrac{1}{N}{\rm E}\big\{\|  \bf{e}^{\hat{\phi}_{\cal C}\to\eta}_t \|^2\}}\big]^{-1},   \\
  \bf{e}^{\eta \to\hat{\phi}_{\cal C}}_t   &\equiv \bf{x}^{\eta \to\hat{\phi}_{\cal C}}_t -\bf{x}, &
  v_t &\equiv \tfrac{1}{N}{\rm E}\big\{\| \bf{e}^{\eta \to\hat{\phi}_{\cal C}}_t  \|^2\big\}.  
\end{align}\ES 
For coded $\bf{x}$, we make the following assumption for OAMP.

\begin{assumption}[Asymptotic IIDG]\label{Pro:SE_new}
Let $\bf{x}\sim \Phi_{\cal C}$ (see \eqref{Eqn:code}). For OAMP in \eqref{OAMP_coded}, $\eta_t$ and $\hat{\phi}_{t}^{\cal C}$ can be characterized by   
 \BE\label{Eqn:TF_C}  
\rho  = {\eta}_{\rm SE}(v),\quad   
v   = \hat{\phi}_{\rm SE}^{\cal C}(\rho)\equiv{\rm mmse}\{\bf{x}|\sqrt{\rho}\bf{x}+\bf{z},\Phi_{\cal C}\},
\EE
where ${\eta}_{\rm SE}(\cdot)$ is the same as that in \eqref{Eqn:TF}.
\end{assumption} 
 
The simulation results in \cite{MaTWC} verify that Assumption \ref{Pro:SE_new} hold for OAMP with convolutional codes (Fig. 4 in \cite{MaTWC}) and LDPC codes (Fig. 7 in \cite{MaTWC}).

\subsection{Curve Matching Principle}

In the un-coded case, OAMP  is not error-free and converges to a non-zero fixed point $(\rho^*,v^*)$.  In the coded case, error-free recovery is possible if $\hat{\phi}_{\rm SE}^{\cal C}(\cdot)$ is properly designed. To obtain an error-free recovery, there should be  no fixed point between $\hat{\phi}_{\rm SE}^{\cal C}(\rho)$ and ${\eta}_{\rm SE}^{-1}({\rho})$, which implies 
\BE\label{Eqn:EF_cond}
    \hat{\phi}_{\rm SE}^{\cal C}(\rho) <  {\eta}_{\rm SE}^{-1}({\rho}),\;\;\; 0 \leq {\rho } \leq snr.
\EE
Also, the decoding MSE should be lower than that of the detector, i.e.,
\BE\label{Eqn:coding_gain}
     \hat{\phi}_{\rm SE}^{\cal C}(\rho) <  \hat{\phi}_{\rm SE}^{\cal S}(\rho),  \;\;\;  {\rm for} \;\; \rho \geq 0.\vspace{-0.1cm}
\EE
Following \eqref{Eqn:EF_cond} and \eqref{Eqn:coding_gain}, OAMP achieves error-free recovery if and only if: For $0 \leq  {\rho } < snr$,
\BE\label{Eqn:upper_bound}
     \hat{\phi}_{\rm SE}^{\cal C}(\rho) < \hat{\phi}_{\rm SE}^*(\rho)\equiv \min\{{\eta}_{\rm SE}^{-1}({\rho}),\; \hat{\phi}_{\rm SE}^{\cal S}(\rho)\}.
\EE 
  
\subsection{Achievable Rate and  Capacity Optimality of OAMP} \label{Sec:area_AMP} 

The lemma below, proved in \cite{Bhattad2007}, establishes the connection between MMSE and code rate. 

\begin{lemma}[Code-Rate-MMSE]\label{Lem:R-MMSE}  Consider a code constraint $\Phi_{\cal C}:\bf{x}\in{\bf{\mathcal{C}}}$. Let the code length be $N$ and code rate $R_{\cal C}$. Then the rate of $\bf{{\mathcal{C}}}$ is given by
\BE\label{Eqn:R_mmse}
R_{\cal C} =  \int_{0}^{\infty}\hat{\phi}_{\rm SE}^{\cal C}(\rho)d\rho =K/N,
\EE
where $\hat{\phi}_{\rm SE}^{\cal C}(\rho) \equiv \tfrac{1}{N}\mr{mmse}\{\bf{x}|\sqrt{\rho}\bf{x}+\bf{z}, \Phi_{\bf{\mathcal{C}}} \}$ is obtained by APP decoding $\eta_{\mathcal C}(\sqrt{\rho}\bf{x}+\bf{z})={\rm E}\{\bf{x}|\sqrt{\rho}\bf{x}+\bf{z}, \Phi_{\bf{\mathcal{C}}}\}$. 
\end{lemma}

Following Lemma \ref{Lem:R-MMSE} and the error-free condition in \eqref{Eqn:upper_bound},  the achievable rate of OAMP receiver is given by
\BE
  R_{\rm OAMP }(snr) = \int_{0}^{\infty} \hat{\phi}_{\rm SE}^{\cal C}(\rho) d\rho,  \quad  \hat{\phi}_{\rm SE}^{\cal C}(\rho)< \hat{\phi}_{\rm SE}^*(\rho). 
\EE
That is, error-free detection can be achieved by the OAMP  receiver if $R_{\cal C}\leq R_{\rm OAMP}$.

\begin{theorem}[Capacity Optimality]\label{The:cap_opt}
 Suppose that Assumption \ref{Pro:SCP}  and  Assumption \ref{Pro:SE_new} hold. Then,  the achievable rate of OAMP achieves the replica constrained capacity of LUIS, i.e.,
\BE\label{Eqn:R_to_C}
R_{\rm OAMP}(snr)\to C^{\rm Rep}_{\rm LUIS}(snr),
\EE 
if $\hat{\phi}_{\rm SE}^{\cal C}(\rho) < \hat{\phi}_{\rm SE}^*(\rho)$ and $\hat{\phi}_{\rm SE}^{\cal C}(\rho) \to \hat{\phi}_{\rm SE}^*(\rho)$ in $[0,  snr]$.  
\end{theorem}

 For a sub-class right-unitarily-invariant $\bf{A}$,  $C^{\rm Rep}_{\rm LUIS}(snr)$ is the true constrained capacity of LUIS \cite{Barbier2018b}. In this case, OAMP is rigorously capacity optimal. Theorem \ref{The:cap_opt} is developed under a matching constraint: $\hat{\phi}_{\rm SE}^{\cal C}(\rho) \to \hat{\phi}_{\rm SE}^*(\rho)$.  The curve-matching code existence is proved in Appendix C-B \cite{LeiTIT} for Gaussian signaling. For non-Gaussian signaling, the curve-matching code existence is still a conjecture.  

 Theorem \ref{The:cap_opt} is based on a matching condition $\hat{\phi}_{\rm SE}^{\cal C}(\rho) 
   \to \hat{\phi}_{\rm SE}^*(\rho)$. An existence proof can be found in Appendix C-B \cite{LeiTIT} for Gaussian signaling. For other signaling, the code existence is a conjecture only. 

\section{Numeric Results}\label{Sec:SIM}

Let the SVD of $\bm{A}$ be $\bm{A}=\bm{U} \bm{\Sigma V}$.  In all the simulations, we set the eigenvalues $\{d_i\}$ in $\bf{\Sigma}$ as \cite{Vila2014}: $d_i/d_{i+1}=\kappa^{1/
T}$ for $i = 1,\ldots, T-1$ and $\sum_{i=1}^Td_i^2=N$, where $T=\min\{M, N\}$. Here, $\kappa\ge1$ controls the condition number of $\bm{A}$. $\bm{U}$ and $\bm{ V}$ are generated by the QR decomposition of two IIDG matrices.
\vspace{-0.2cm}

\begin{figure}[t] 
  \centering
  \includegraphics[width=6cm]{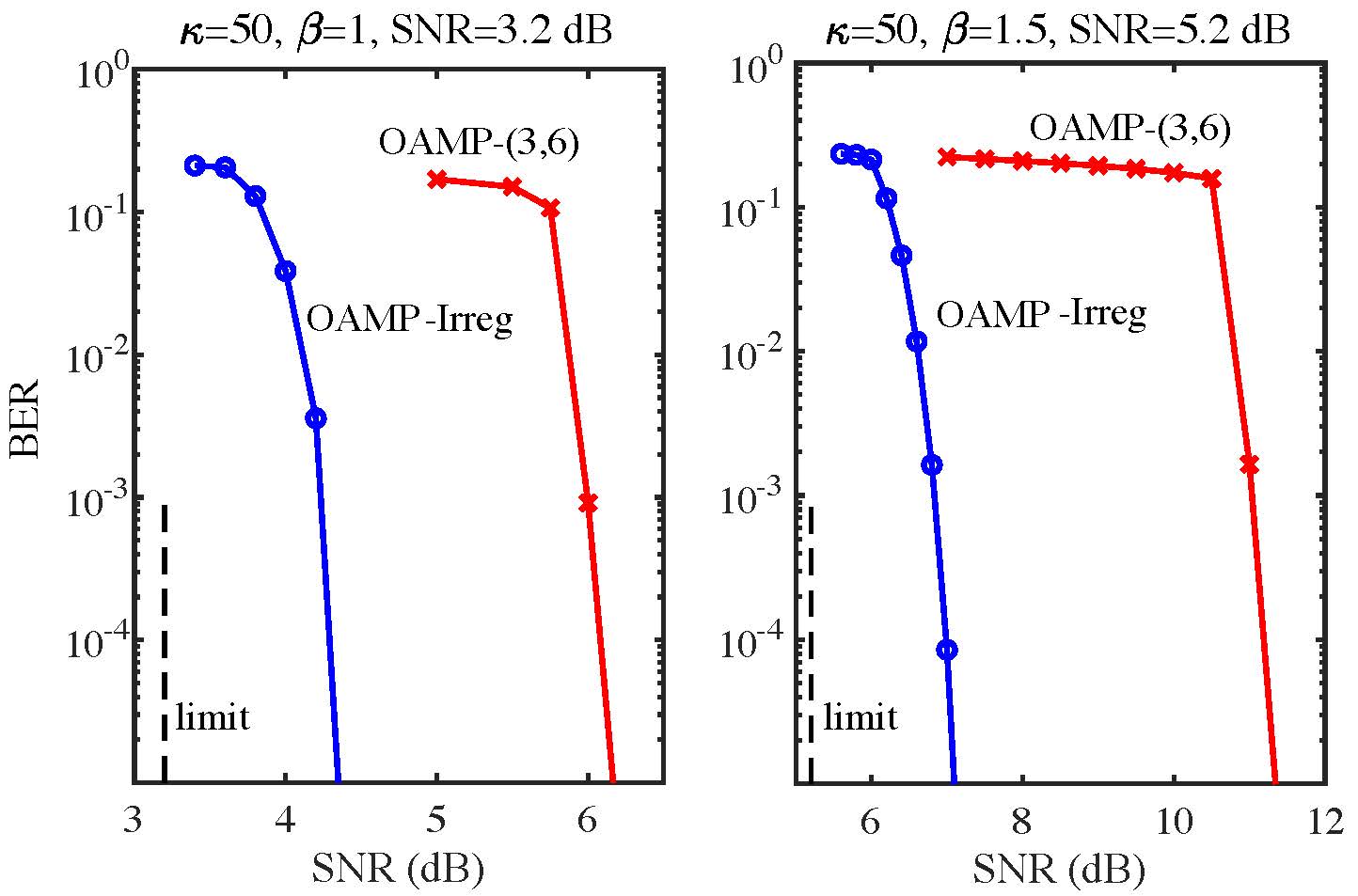}\\ \vspace{-0.2cm}
  \caption{ BER performances of OAMP , where ``OAMP-Irreg'' denotes the BER of optimized irregular LDPC codes, ``OAMP-(3, 6)'' the BER with regular (3, 6) LDPC codes. Codeword length = $10^5$, code rate $=$ 0.5075 (left) and 0.4721 (right), QPSK modulation, and iterations = $250$, and $\beta=N/M=\{1, 1.5\}$.}\label{Fig:BER_OAMP} \vspace{-0.2cm}
\end{figure} 
  
Fig. \ref{Fig:BER_OAMP} shows the BER simulations for the LUIS, where $\bf{x}$ is optimized irregular LDPC codes \cite{Yuan2008Low, Chung2001}. The receiver ``OAMP-Irreg'' is shown in Fig. \ref{Fig:OAMP}, and the NLE is a sum-product decoder. The system sizes are $(N, M)=(500, 500) \;\mr{and}\; (500, 333)$. At BER $=10^{-5}$, the curves of ``OAMP-Irreg''  are about $1$~dB away from their limits. We compare the ``OAMP-Irreg''  with ``OAMP-(3, 6)'', i.e., the cascading OAMP with un-optimized regular (3, 6) LDPC codes \cite{Gallager1962}. At  BER = $10^{-5}$, the proposed ``OAMP-Irreg''  outperforms ($0.8 \sim 4$~dB gains) ``OAMP-(3, 6)'' for  $\beta=\{1, 1.5\}$ and $\kappa=\{10, 50\}$. In other words, code optimization can bring significant performance improvement for OAMP.

\begin{figure}[t] \vspace{-0.2cm}
  \centering
  \includegraphics[width=6cm]{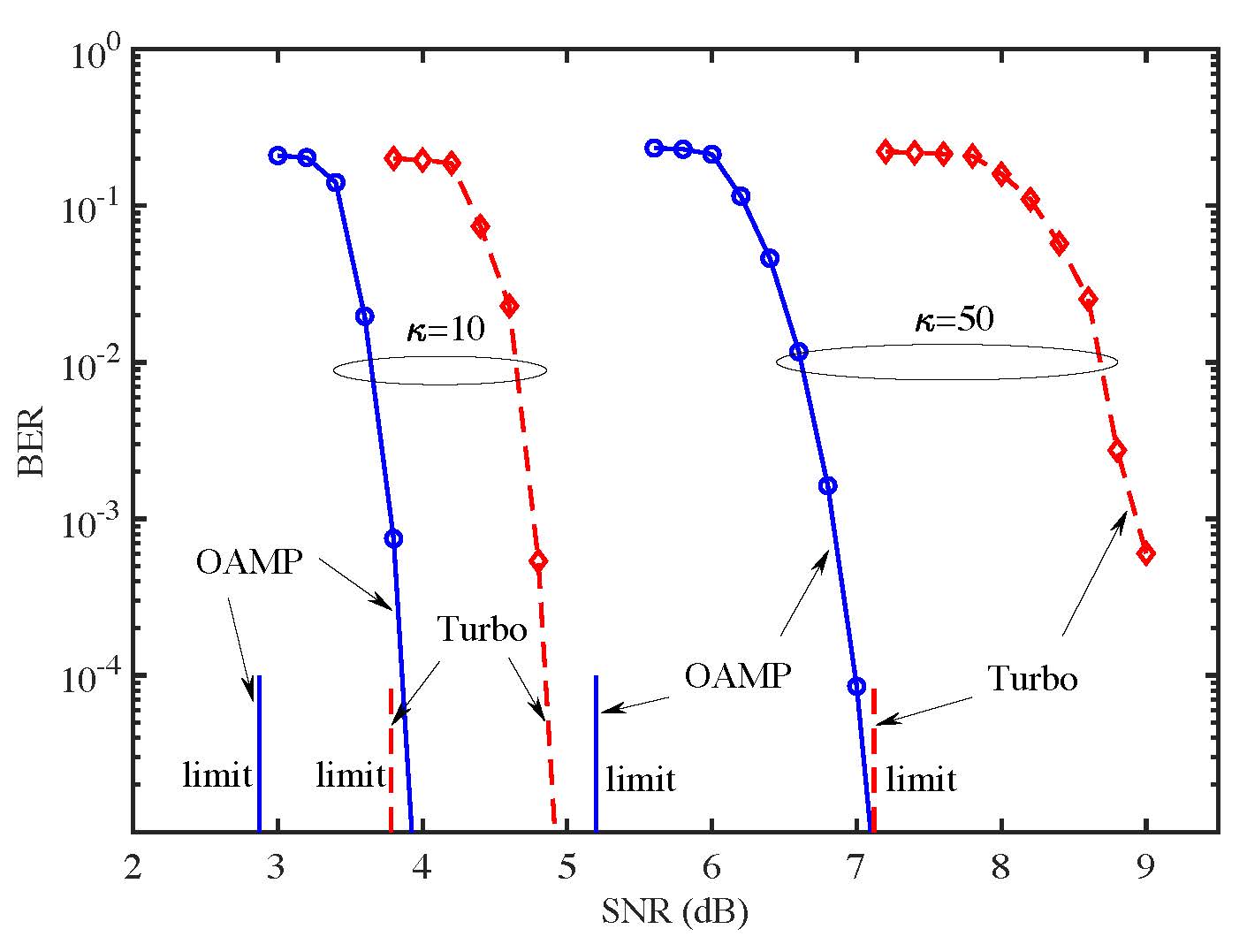}\\\vspace{-0.2cm}
  \caption{ BER performances of OAMP and Turbo   \cite{YC2018TWC, Lei20161b} with optimized irregular LDPC codes, where  ``limit" denotes the QPSK rate limits of OAMP or Turbo. Codeword length = $10^5$, code rates $=\{0.5062, 0.4721, 0.5008, 0.4863\}$ (for the curves from left to right), QPSK modulation, and iterations = $250$, and $\beta=1.5$ with $N=500$ and $M=333$. }\label{Fig:BER_OAMP_Turbo} \vspace{-0.2cm}
\end{figure}

Fig.~\ref{Fig:BER_OAMP_Turbo} compares OAMP with the conventional Turbo \cite{ YC2018TWC, Lei20161b}. A QPSK modulated LUIS with $(N, M)=(500, 333)$ and $\kappa=\{10, 50\}$ (condition number) is considered.  Fig.~\ref{Fig:BER_OAMP_Turbo} compares the BERs of the optimized OAMP and the optimized Turbo (with iterations $=250$). The simulated BERs of OAMP and Turbo are about 1dB for $\kappa=10$ and 2 dB for $\kappa=50$ away from their limits respectively. Comparing with the Turbo, OAMP has 1 dB improvement for $\kappa=10$ and 2 dB improvement for $\kappa=50$ in BER. Overall, the conventional Turbo has huge performance loss in general discrete linear systems, especially in the case of high transmission rate and high conditional number.

\section{Conclusion}
An OAMP receiver is considered for a coded LUIS with a unitarily invariant sensing matrix and an arbitrary input distribution. Several area properties are established using the established properties of OAMP. We show that OAMP is capacity optimal under the assumptions that the SE for joint OAMP and FEC decoding is correct and the replica method is reliable. A curve-matching coding principle is developed for OAMP. Simulation results are provided to verify that the OAMP with optimized irregular LDPC codes approaches the replica capacity of LUIS, and significantly outperforms ($2$ dB $\sim$ $4$ dB gain) the un-optimized case. Furthermore, the OAMP has a significant improvement in BER performance over the state-of-art Turbo-LMMSE.

\end{document}